\documentclass[journal]{IEEEtran}

\usepackage{graphicx}
\usepackage[cmex10]{amsmath}
\usepackage{array}
\usepackage{url}
\usepackage{amsfonts}
\usepackage{amssymb}
\usepackage{algorithm}
\usepackage{algorithmic}
\usepackage{color}

\newcommand{\ket}[1]{|{#1}\rangle}

\newcommand{\beq}{\begin{equation}}
\newcommand{\eeq}{\end{equation}}

\def\<{\langle}
\def\>{\rangle}

\newtheorem{definition}{Definition}
\newtheorem{example}{Example}

\newtheorem{theorem}{Theorem}
\newtheorem{corollary}{Corollary}

\newtheorem{fact}{Fact}

\begin{document}

\title{Codeword Stabilized Quantum Codes:\\ Algorithm $\&$ Structure}

\author{Isaac~L.~Chuang, 
Andrew~W.~Cross, 
Graeme~Smith, 
John~Smolin, 
        and~Bei~Zeng 
}

\maketitle

\begin{abstract}
The codeword stabilized (``CWS'') quantum codes formalism presents a
unifying approach to both additive and nonadditive quantum
error-correcting codes (arXiv:quant-ph/0708.1021).  This formalism
reduces the problem of constructing such quantum codes to finding a
binary classical code correcting an error pattern induced by a graph
state.  Finding such a classical code can be very difficult.  Here, we
consider an algorithm which maps the search for CWS codes to a problem
of identifying maximum cliques in a graph. While solving this problem
is in general very hard, we provide three structure theorems which
reduce the search space, specifying certain admissible
and optimal $((n,K,d))$ additive codes.  In particular, we find there
does not exist any $((7,3,3))$ CWS code though the linear programing
bound does not rule it out.
The complexity of the CWS-search algorithm is compared with
the contrasting method introduced by Aggarwal $\&$ Calderbank
(arXiv:cs/0610159).
\end{abstract}

\section{Introduction}
Quantum error correcting codes play a significant role in quantum
computation and quantum information.  While considerable understanding
has now been obtained for a broad class of quantum codes, almost all
of this has focused on stabilizer codes, the quantum analogues of
classical additive codes.  Recently, a number of {\em nonadditive}
quantum codes have been discovered, with superior coding parameters
$((n,K,d))$, the number of physical qubits being $n$, the dimension of
the encoded space $K$, and the code distance $d$ \cite{CSSZ:07,Yu:07a,Yu:07b}.
These new codes have inspired a search for more high-performance
non-additive quantum codes \cite{Grassl:08b}, a desire to understand how
non-additive codes relate to additive codes, and how these may be
understood through a cohesive set of basic principles.

A systematic construction, providing a unifying approach to both
additive and nonadditive quantum error-correcting codes, has been
obtained \cite{CSSZ:07}.  This {\em codeword stabilized quantum codes}
(``CWS'' quantum codes) approach constructs the desired quantum code
based on a binary classical code $\mathcal{C}$, chosen to correct a
certain error pattern induced by a self-dual additive quantum code
which is without loss of generality, taken to be a graph state
$\mathcal{G}$.  The construction thus reduces the problem of finding a
quantum code into a problem of finding a certain classical code.  All
previously known nonadditive codes \cite{Rains:97a,Smolin:07a,Yu:07a,Feng:08}
with good parameters can be constructed within the CWS construction.

\begin{figure}[htbp]
\centering
\includegraphics[width=2.00in]{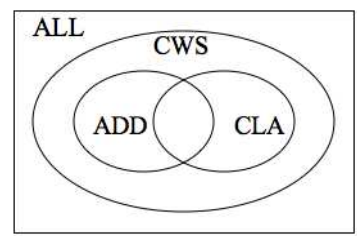}
\caption{The relationship of CWS codes with additive quantum codes and
classical codes: ALL: all quantum codes; CWS: CWS codes; ADD: additive
codes; CLA: classical codes.} \label{fig1}
\end{figure}

The natural challenge in these approaches is efficient identification
of suitable classical codes, from which the desired additive and
non-additive quantum codes can be constructed.  It is apparent that
due to the error pattern induced by the graph state $\mathcal{G}$, the
binary classical code $\mathcal{C}$ does not coincide with the usual
binary classical code where the minimum Hamming distance is a more
important code parameter -- although interestingly, they do coincide in
the special case where $\mathcal{G}$ is an unconnected graph, so the
family of CWS quantum codes includes classical (``bit-flip'') codes
as depicted in Fig. \ref{fig1}.

The CWS construction, observing that a classical code correcting
certain bit-flip error patterns gives rise to a quantum code, allows a
natural encoding of the problem of finding a quantum code
$\mathcal{Q}=(\mathcal{G},\mathcal{C})$ into an equivalent problem, of finding the
maximum clique of an induced graph, called the CWS clique graph.  The
existence of such a mapping is not surprising, since {\sc maxclique}
is an NP-complete problem \cite{Sipser:05,Garey-Johnson:79}, and thus
can be used for a reduction from all unstructured search problems.  In
practice, many heuristic and randomized Clique solvers and SAT solvers
have been developed, with reasonable run-times for small problem
sizes.  And since the search for CWS codes starts from a graph state
$\mathcal{G}$, prior art in categorizing local Clifford (LC) orbits of
those states \cite{Danielsen:05a,Danielsen:05b} helps simplify the
problem.  Nevertheless, without further simplification, a mapping of
the CWS quantum codes search problem to {\sc maxclique} leaves the
problem unsolved, due to the exponential computational cost of solving
{\sc maxclique}. The real situation is even worse. For a general graph
state, the search problem is NP-complete due to the reduction to {\sc
maxclique}. However, to search for all the quantum codes, we need to
search for all graphs of $n$ vertices, which contributes a factor of
order $2^{n^2}$.

Here, we present an algorithm for finding CWS codes, based on a
mapping to {\sc maxclique}.  We show that despite the exponential
complexity of solving this {\sc cws-maxclique} problem, the algorithm
can be usefully employed to locate and identify a wide variety of
codes, by taking careful steps to prune the search space.  In
particular, we show how the complexity cost can be reduced by using
known graph isomorphisms and LC equivalences of graph states.  We also
present simplifying criteria for the search, arising from the
structural properties of CWS codes.  We prove three theorems limiting
whether $((n,K,d))$ additive codes with optimal $K$ can be improved,
or not, by the CWS construction.  These theorems allow significant
practical reduction of the search space involved in finding CWS codes
using {\sc cws-maxclique}.  Furthermore,
these theorems also indicate the existence of quantum codes outside of
the CWS construction, as alluded to in Fig.~\ref{fig1}.

We also compare and contrast the CWS codes with another
framework (``AC06'') which was introduced
independently \cite{Calderbank:06a} and is based on a correspondence
between Boolean functions and projection operators.  
We interpret the AC06 framework to use a quantum state and a
classical code, to generate the desired quantum code, but in a
sense, it works in the reverse direction, starting from the classical
code and obtaining the quantum state.
We show how the AC06
Boolean function $f$ is the analogue of our classical code
$\mathcal{C}$, up to a LC equivalence.  This allows us to extend AC06
to degenerate codes, and to show that the AC06 framework can also be
used to construct a search algorithm for new quantum codes, with
comparable complexity to {\sc cws-maxclique}.

\section{The {\sc cws-maxclique} algorithm}
\label{sec:alg}

The {\sc cws-maxclique} algorithm is a procedure to search for a
quantum error correction code $\mathcal{Q}=(\mathcal{G},\mathcal{C})$,
given a graph state $\mathcal{G}$ which maps quantum errors ${\mathcal
E}$ in the Pauli group
into binary error patterns, and a classical code ${\mathcal{C}}$,
which corrects the error patterns.  We present this algorithm below,
beginning with a review of the basic definitions of CWS codes,
proceeding to the details of the procedure, then rounding up with an
evaluation of the computational complexity of the algorithm.

\subsection{Non-degenerate and degenerate CWS codes}

The basic concepts and definitions of CWS codes are described in a
previous paper\cite{CSSZ:07}, and may be summarized as follows.  The
{\bf standard form CWS code} is fully characterized by a graph
$\mathcal{G}$ and a classical binary code $\mathcal{C}$, such that the
corresponding CWS code may be denoted by the pair ${\mathcal Q} =
(\mathcal{G},\mathcal{C})$.  We define
\begin{equation}
	Cl_{\mathcal{G}}({\mathcal E})=\{ Cl_{\mathcal{G}}(E)\ |\ E\in
	{\mathcal E}\}
\end{equation}
as the set of classical errors induced by quantum errors ${\mathcal
E}$ acting on the graph $\mathcal{G}$; these are the errors that the
classical code $\mathcal{C}$ must detect.  For each quantum error $E$,
it is sufficient to express $E$ in Pauli form as 
$E=\pm Z^{\mathbf v}X^{\mathbf u}$ for some bit
strings ${\mathbf u}$ and ${\mathbf v}$. The mapping to classical
error strings is
\begin{equation}
	Cl_\mathcal{G}(E=\pm Z^{\mathbf v}X^{\mathbf u})={\mathbf v}\oplus \bigoplus_{l=1}^n ({\mathbf u})_l{\mathbf r}_l\label{pattern}
\,,
\end{equation}
where $\mathbf{r}_l$ is the $l$th row of the adjacency matrix for
$\mathcal{G}$, and $(\mathbf{u})_l$ is the $l^{th}$ bit of $\mathbf{u}$.  

Using these definitions, the main theorem of the CWS code construction
(Theorem 3 of \cite{CSSZ:07}) may be given as:
\begin{theorem}
\label{CSSZTheorem3}
A standard form CWS code, ${\mathcal Q} = (\mathcal{G},\mathcal{C})$
for graph state $\mathcal{G}$ and classical code ${\mathcal{C}}$,
detects errors from $\mathcal{E}$ if and only if $\mathcal{C}$ detects
errors from $Cl_\mathcal{G}(\mathcal{E})$ and in addition, for each $E
\in {\mathcal E}$,
\begin{eqnarray}
\label{clseneq0}{\rm either}~~ Cl_\mathcal{G}(E) & \neq&  0\\
\label{complus} {~\rm or ~~} \forall i\ Z^{{\mathbf c}_i}E&=&EZ^{{\mathbf c}_i}
\,,
\end{eqnarray}
where $Z^{{\mathbf c}_i}$ are codeword operators for ${\mathcal{C}}$
from $\{ Z^{\mathbf c}\}_{{\mathbf c} \in \mathcal{C}}$.
\end{theorem}

The case where $Cl_{\cal G}(E)\neq 0$ for all $E\in {\cal E}$ is the 
non-degenerate case. For 
degenerate CWS codes, it will be useful to introduce a new set of
classical bitstrings
\begin{align}
	D_\mathcal{G}({\mathcal E})
	=\{ & \mathbf{c}\in \{0,1\}^n\ |\ Cl_{\cal G}(E)=0\ \textrm{and}\ \\
	& {\mathbf c}\cdot{\mathbf u}\neq 0\ \textrm{for some}\ E=\pm Z^{\mathbf v}X^{\mathbf u}\in {\mathcal E}\}
\,.
\end{align}
These bitstrings indicate codewords which are inadmissible, because they
violate the condition given by equations (\ref{clseneq0}) 
and (\ref{complus}) of Theorem~\ref{CSSZTheorem3}. 
Specifically, fix a codeword ${\bf c}$, then for all $E\in {\cal E}$
we must have $Z^{\bf c}E=EZ^{\bf c}$ if $Cl_{\cal G}(E)=0$.
Writing $E=\pm Z^{\mathbf v}X^{\mathbf u}$, ${\bf c}$ is not an
admissible codeword if $Cl_{\cal G}(E)=0$ and 
${\mathbf c}\cdot {\mathbf u}\neq 0$.
In other words, if a CWS code is degenerate, some low weight errors act 
trivially on the code space (i.e. $Cl_{\cal G}(E)=0$), and these errors 
must act trivially on each basis state generated from the graph state 
${\mathcal G}$ (i.e. $[Z^{\mathbf c},E]=0$). 
$D_\mathcal{G}({\mathcal E})$ describes basis states for which this is not 
the case.

\subsection{The {\sc cws-maxclique} algorithm}

Given a graph ${\mathcal G}$, the problem of finding a CWS code
$\mathcal{Q}=(\mathcal{G},\mathcal{C})$, which corrects for quantum
errors ${\cal E}$, is reduced to a search for suitable classical
codes.  It is thus natural to ask how such classical codes can be
found.  One solution might be to use existing classical codes for this
construction.  However, that approach gives sub-optimal code
parameters, due to the fact that ${\mathcal{C}}$ should be able to
detect errors of the highest weight of the induced error patterns in
$Cl_\mathcal{G}({\mathcal E})$.  This means that the classical code
${\mathcal{C}}$ must have distance significantly greater than that of
the corresponding quantum code $(\mathcal{G},\mathcal{C})$, as shown
in the following example:

\begin{example}
Let $\mathcal{G}$ be an $n$ qubit ring graph.  If ${\mathcal{E}}$ is
the set of single qubit Pauli $X$, $Y$, and $Z$ errors, then the
induced classical errors $Cl_\mathcal{G}({\mathcal E})$ are single,
triple, and double bit flips respectively.  Choosing the classical
code $\mathcal{C}$ to be a binary $((n,K,7))$ code results in a CWS
code $(\mathcal{G},\mathcal{C})$ with parameters $((n,K,3))$.
However, $\mathcal{C}$ also detects many additional errors which are
unnecessary for this construction, such as all the one to six bit flip
errors; $Cl_\mathcal{G}({\mathcal E})$ only includes a subset of those
errors.
\label{classicalcode}
\end{example}

This example motivates a search for specific classical codes which
correct just the relevent errors for the CWS construction.  However,
classical coding theory provides no efficient, systematic
constructions for codes that correct the potentially exotic error
patterns involved in the CWS construction.  On the other hand, finding
a code with the best $K$ for given $n$ and $d$ is a problem which can
be naturally encoded into an NP-complete problem such as {\sc
maxclique}.  This classic approach has been employed, for example, to
show that the $(10,K,3)$ classical code with $K=72$ has optimal
parameters\cite{Ostergard:99a}.

{\sc cws-maxclique} is a mapping onto {\sc maxclique}, of the problem
of finding the CWS code $(\mathcal{G},\mathcal{C})$ with the largest
possible dimension $K$, for given parameters $n$, $d$, and graph
$\mathcal{G}$.  The {\sc cws-maxclique} algorithm gives steps to solve
this problem, and is given in detail in the
Algorithm~\ref{alg:CWSMaxClique} box.  It proceeds in several simple
steps.  The first step, \textbf{Setup}$({\mathcal E},\Lambda)$
(Algorithm~\ref{alg:setup}), finds the elements of
$Cl_\mathcal{G}({\mathcal E})$ and $D_\mathcal{G}({\mathcal E})$. The
second step, \textbf{MakeCWSCliqueGraph}$(\textsc{CL},\textsc{D})$
(Algorithm~\ref{alg:MakeCWSCliqueGraph}), constructs a graph, denoted
as the CWS ``clique graph,'' whose vertices are classical codewords
and whose edges indicate codewords that can be in the same classical
code together.  When searching for ordinary classical codes using an
analogous procedure, the usual condition for joining two vertices by
an edge is that the vertices are Hamming distance $d$ apart. In our
situation, vertices are joined by an edge if there is no error induced
by the graph state that maps one codeword to the other.  Finally, an
external subroutine
\textbf{findMaxClique}$(V,E)$ is called; this routine is to employ known
techniques to find the maximum clique in the CWS clique graph. The
clique-finding subroutine is not specified here because there are many
exact and heuristic techniques known in the community, for solving
this classic NP-complete problem.  Note that in the detailed
description of the algorithms, two functions are used:
$\text{String}(i):\ \text{integer}\ i \rightarrow \text{binary string
of}\ i\ \text{with length}\ n$, and its inverse, $\text{Integer}(i):\
\text{binary string \text{with length}\ n}\ i \rightarrow
\text{integer of}\ i$. Also, an error configuration is a list of ordered
pairs $(\textsc{LOC},\textsc{TYPE})$ where $\textsc{LOC}$ is the coordinate of the
affected qubit and $\textsc{TYPE}$ is one of $X$, $Y$, or $Z$.

\begin{algorithm}
\caption{\textbf{Setup}$({\mathcal E},\Lambda)$: Compute $Cl_{\mathcal G}({\mathcal E})$ and $D_{\mathcal G}({\mathcal E})$, where $\mathcal
E$ is a set of Pauli errors
and $\Lambda$ is the adjacency matrix associated with graph
$\mathcal{G}$.}
\label{alg:setup}
\algsetup{indent=2em}
\begin{algorithmic}[1]
\REQUIRE $\Lambda^T=\Lambda$, $\Lambda_{ij}=\{0,1\}$ and
$\Lambda_{ii}=0$ \ENSURE \textsc{CL}$[i]=\delta(\text{String}(i)\in Cl_{\mathcal{G}}({\mathcal
E}))$ and $\textsc{D}[i]=\delta(\text{String}(i)\in D_{\mathcal{G}}({\mathcal E}))$
\FOR{$i\in\{0,1\}^n$} \STATE $\textsc{CL}[\text{Integer}(i)]\leftarrow 0$ \STATE
$\textsc{D}[\text{Integer}(i)]\leftarrow 0$ \ENDFOR 
\FOR{error configuration $E\in {\mathcal E}$} 
\STATE \textsc{err}$\leftarrow \text{String}(0)$ \STATE
\textsc{errx}$\leftarrow \text{String}(0)$ \FOR{$(\textsc{loc},\textsc{type})$ in
$E$} \IF{\textsc{type} is X or Y} \STATE \textsc{err} $\leftarrow$
\textsc{err} $\oplus\ (\text{row}\ \textsc{loc}\ \text{of}\ \Lambda)$ \STATE \textsc{errx}
$\leftarrow$ \textsc{err} $\oplus\ \text{String}(2^{\textsc{loc}})$ \ENDIF
\IF{\textsc{type} is Z or Y} \STATE \textsc{err} $\leftarrow$
\textsc{err} $\oplus\ \text{String}(2^{\textsc{loc}})$ \ENDIF \ENDFOR \STATE
\textsc{CL}[\text{Integer}(\textsc{err})] $\leftarrow 1$ 
\IF{\text{Integer}(\textsc{err}) is $0$}
\FOR{$i\in\{0,1\}^n$} \IF{$\textsc{errx}\cdot i\neq 0$} \STATE
\textsc{D}[i] $\leftarrow 1$ \ENDIF \ENDFOR \ENDIF \ENDFOR \RETURN
$(\textsc{CL},\textsc{D})$
\end{algorithmic}
\end{algorithm}

\begin{algorithm}
\caption{\textbf{MakeCWSCliqueGraph}$(\textsc{CL},\textsc{D})$:
Construct a graph whose vertices $V$ are classical codewords and whose
edges $E$ connect codewords that can belong to the same classical code,
according to the error model indicated by $Cl_{\mathcal G}({\mathcal
E})$ and $D_{\mathcal G}({\mathcal E})$.}
\label{alg:MakeCWSCliqueGraph}
\algsetup{indent=2em}
\begin{algorithmic}[1]
\REQUIRE $\textsc{CL}$ and $\textsc{D}$ are binary arrays of length $2^n$
\ENSURE $0^n\in V$, $0^n\neq v\in V\Rightarrow \textsc{D}[v]=0$ and $\textsc{CL}[v]=0$, $(v,w)\in E\Rightarrow \textsc{CL}[v\oplus w]=0$
\STATE $V\leftarrow \{0^n\}$
\STATE $E\leftarrow\emptyset$
\FOR{$s\in \{0,1\}^n$}
\IF{\textsc{D}$[s]=0$ and $\textsc{CL}[s]=0$}
\STATE $V\leftarrow V\cup\{s\}$
\FOR{$v\in V\setminus\{s\}$}
\IF{\textsc{CL}$[v\oplus s]=0$}
\STATE $E\leftarrow E\cup\{(v,s)\}$
\ENDIF
\ENDFOR
\ENDIF
\ENDFOR
\RETURN $(V,E)$
\end{algorithmic}
\end{algorithm}

\begin{algorithm}
\caption{\textbf{CWS-MAXCLIQUE}$({\mathcal E},\Lambda)$: Find a quantum code 
$\mathcal{Q}$ detecting errors in $\mathcal{E}$, and providing the
largest possible dimension $K$ for the given input.  The input
$\Lambda$ specifies the adjacency matrix of the graph ${\mathcal G}$.
The output ${\mathcal C}$ is a classical code such that ${\mathcal
Q}=(\mathcal{G},\mathcal{C})$ is a CWS code detecting errors in
$\mathcal{E}$.}
\label{alg:CWSMaxClique}
\algsetup{indent=2em}
\begin{algorithmic}[1]
\REQUIRE  $\Lambda^T=\Lambda$, $\Lambda_{ij}=\{0,1\}$ and $\Lambda_{ii}=0\ \forall i$
\ENSURE $K=|\mathcal{C}|$ is as large as possible for the given input,
$0^n\in\mathcal{C}$, and
$\mathcal{C}$ satisfies the conditions in Theorem 3 of \cite{CSSZ:07} 
\STATE $(\textsc{CL},\textsc{D})\leftarrow \textbf{Setup}({\mathcal E},\Lambda)$
\STATE $(V,E)\leftarrow \textbf{MakeCWSCliqueGraph}(\textsc{CL},\textsc{D})$
\STATE $\mathcal{C}\leftarrow \textbf{findMaxClique}(V,E)$
\RETURN $\mathcal{C}$
\end{algorithmic}
\end{algorithm}

\subsection{The complexity}

{\sc cws-maxclique} is not an efficient algorithm; the run-time is at
least of order $\sim 2^n$, because of the representation of the
bit-string sets $Cl_{\mathcal{G}}({\mathcal E})$ and
$D_{\mathcal{G}}({\mathcal E})$.  These are needed to specify the CWS
clique graph, which has $2^n$ nodes.  In principle, instead of storing
all this in memory, the vertices and edges of this graph could be
computed on the fly, during execution of the {\bf findMaxClique}
subroutine.  However, these inefficiencies are not limiting factors,
because of the even larger size of the search space involved in
typical applications.

Typically, the goal is not to search for an optimal CWS code, given
${\mathcal G}$ and ${\mathcal E}$, but rather, to determine if an $((n,K,d))$ code exists when $n$ and $K$ are fixed.
When $K$ is fixed, finding a maximum clique is not
necessary; rather, a clique of size $K$ is desired.  There are ${2^n
\choose K}$ such possible cliques.  Checking whether a size $K$
subgraph of a CWS clique graph is a clique just requires checking if
that subgraph is fully connected.  Given an adjacency matrix for the
CWS clique graph (and constant time access to the matrix elements),
checking a subgraph takes order $K^2$ steps.

Searching over the space of all possible graphs ${\mathcal G}$ involves
searching a space of graphs with $n$ vertices, with a total of $2^{n
\choose 2}$ possibilities. Therefore, the complexity of searching for an
$((n,K,d))$ CWS code is roughly
\begin{equation}
	K^2 2^{n \choose 2}{2^n \choose K}.\label{complexity}
\end{equation}

However, several practical improvements allow this search space to be
pruned usefully.  First, not all graphs ${\mathcal G}$ need be
considered; only those which are inequivalent under local Clifford
(LC) operations need be checked.  The LC orbits of graphs are well
understood, and efficient algorithms exist to check for LC equivalence
\cite{Danielsen:05a,Danielsen:05b,VanDenNest:04a}. Therefore, the
factor $2^{n \choose 2}$ can be significantly reduced. A lower bound
on the number of LC inequivalent graphs is given in
\cite{Bahramgiri:06}, based on the number of non-isomorphic tree
graphs, which roughly scales as $3^{n}$.  This reduction has played a
key role in allowing us to employ the {\sc cws-maxclique} algorithm on
spaces with parameters up to $n=11$ and $K=32$.  However, no suitable
upper bound is presently known, which would give a quantitative
estimate of the extent of the search space reduction due to LC
equivalence.

A second practical improvement comes from intrinsic properties of CWS
codes, which rule out existence of codes of certain $((n,K,d))$
parameters, and relate the existence of certain parameter values with
the existence of others.  We will return to discuss these structure
theorems in Section~\ref{sec:struct}.

\section{Boolean functions and Classical Codes}
\label{sec:bfunc}

The CWS construction unifies all known additive and non-additive
quantum error correction codes of good parameters, including both
degenerate and non-degenerate codes.  An alternative
framework (``AC06'') for non-degenerate codes, has been presented by
Aggarwal $\&$ Calderbank \cite{Calderbank:06a}, based on a
correspondence between Boolean functions and projection operators.
Because AC06 implies a search algorithm for quantum codes which is in
a sense the reverse of that employed above, in {\sc cws-maxclique}, it
is interesting to consider the differences.

In this section we study the relationship between AC06 and the CWS
construction, by linking the AC06 Boolean function, which we interpret
to specify a certain classical code, to the classical code $\mathcal{C}$ 
used in the CWS construction. The components of the AC06 construction can
be naturally associated with those of the CWS construction. In this way, 
we show that AC06 codes are spanned by a set of stabilizer states 
generated from a single state and a set of Pauli operators. Therefore,
AC06 codes can be described completely, and in our opinion more transparently, 
as CWS codes.

That this identification between AC06 and CWS is natural was mentioned 
previously
\cite{CSSZ:07}, but the transform required has not been presented before.
It is well known that any stabilizer state is
equivalent under some LC transform to a graph state.  Thus, supposing
that a local Clifford operation maps the AC06 stabilizer state to a graph 
state, it would be nice if this Clifford also described
a transform from the Boolean function $f$ to the binary classical code
${\mathcal C}$ of the CWS construction.
Below, we show this mapping indeed exists, up to a technical subtlety
with regard to the choice of the generating set for the stabilizer.

The AC06 framework is not entirely complete since degenerate codes cannot
be described as presented in \cite{Calderbank:06a}.
Degenerate codes may, in some cases, outperform the best known nondegenerate
codes. Such an example may be provided by the $[[25,1,9]]$ code obtained by
concatenating the $[[5,1,3]]$ code, since this is the best known $[[25,1]]$
code, it is degenerate, there is no known nondegenerate $[[25,1,9]]$, and 
it has the highest possible minimum distance \cite{Grassl:tables}.
We take the constraints given for degenerate codes in the CWS
construction and map these backwards to given new constraints for 
degenerate codes in the AC06 framework.

Given a complete AC06 framework which includes both
non-degenerate and degenerate codes, we can then compare and contrast
the computational cost of the CWS and AC06 approaches for seeking
optimal parameter quantum codes.
When the search goal is to find an optimal 
$((n,K,d))$ code for fixed $n$ and $K$, the AC06 framework seems at first to
involve a search over possibly $2^{2^n}$ Boolean functions, while {\sc
cws-maxclique} involves a search over $2^{n \choose 2}$ possible
graphs.  This appears to give significant advantage to {\sc
cws-maxclique}.  However, we find that with careful analysis of AC06,
and extending it include degenerate codes, the two 
search algorithms have comparable complexity.

\subsection{AC06 quantum error-correcting codes are CWS codes}

A $n$-variable Boolean function is a mapping $f:\{0,1\}^n\rightarrow \{0,1\}$ that maps a binary $n$-vector
${\bf v}=(v_1,\dots,v_n)$ to a bit $f(v_1,\dots,v_n)$. A Boolean function is nonzero if there exists some 
${\bf v}$ such that $f({\bf v})=1$. We know that a Boolean function is naturally associated with a classical code 
\beq
{\cal C}_f=\{ {\bf c}\in\{0,1\}^n\ |\ f({\bf c})=1\}.
\eeq
A nonzero Boolean function $f$ can be represented as 
\beq
f({\bf v})=\sum_{{\bf c}\in {\cal C}_f} v_1^{c_1}v_2^{c_2}\dots v_n^{c_n},
\eeq
where $v_i^1=v_i$ and $v_i^0=\bar{v_i}=v_i\oplus 1$. The summation is taken to be modulo $2$, i.e. XOR. 
The weight of a Boolean function $f$ is $|{\cal C}_f|$.

The complementary set of a nonzero $n$-variable Boolean function $f({\bf v})$ is defined by
\beq
Cset_f=\{{\bf a}\in \{0,1\}^n\ |\ \sum_{{\bf c}\in {\cal C}_f} f({\bf c})f({\bf c}\oplus {\bf a})=0\}.
\eeq
We know that the complementarly set is simply the set of vectors ${\bf a}$ such that 
${\cal C}_f\cap ({\cal C}_f\oplus {\bf a})=\emptyset$, i.e. it is the set of (classical) detectable errors of
${\cal C}_f$, since no codeword is mapped back into the code by ${\bf a}$.

\begin{definition}[Definition 6 of \cite{Calderbank:06a}]
Let $P$ and $Q$ be projection operators on a Hilbert space $H$ with $K=\textrm{image}(P)$ and
$L=\textrm{image}(Q)$. Then
\begin{itemize}
\item $P<Q$ iff $K\subset L$ and $K\neq L$
\item $P\vee Q$ is the projection of $H$ onto the span $K\vee L$, the smallest subspace of $H$ containing both $K$ and $L$
\item $P\wedge Q$ is the projection of $H$ onto $K\cap L$
\item $\bar{P}$ is the projection of $H$ onto $K^\perp$
\item $P\oplus Q=(P\wedge\bar{Q})\vee(\bar{P}\wedge Q)$.
\end{itemize}
\end{definition}

\begin{definition}[Definition 7 of \cite{Calderbank:06a}]
Given an arbitrary Boolean function $f(v_1,\dots,v_n)$, the projection function
$f(P_1,P_2,\dots,P_n)$ is the expression in which $v_i$ in the Boolean function is replaced by the projection
operator $P_i$, multiplication
(AND) in the Boolean logic is replaced by the meet operation $P\vee Q$ in the projection logic, summation (OR) 
in the Boolean logic is replaced by the join operation $P\wedge Q$ in the projection logic, and the NOT
operation in the Boolean logic is replaced by the not operation $\bar{P}$ in the projection logic.
Note that summation modulo $2$ (XOR) is replaced by the cooresponding operation $P\oplus Q$ in the
projection logic.
\end{definition}

\begin{theorem}[Theorem 1 of \cite{Calderbank:06a}]
If $(P_1,P_2,\dots,P_n)$ are pairwise commutative projection operators of dimension $2^{n-1}$ such that
$(P_1P_2\dots P_n)$, $(P_1P_2\dots \bar{P_n})$, \dots, $(\bar{P_1}\bar{P_2}\dots\bar{P_n})$ are all one-dimensional
projection operators and $H$ is of dimension $2^n$, then $P_f=f(P_1,P_2,\dots,P_n)$ is an orthogonal projection
on a subspace of dimension $K=\textrm{Tr}(P_f)=\textrm{wt}(f)$.
\end{theorem}

Let $({\bf a}|{\bf b})$ denote the concatenation of two $n$-bit binary vectors ${\bf a}$ and ${\bf b}$. 
The symplectic inner product of $2n$-bit binary vectors $({\bf a}|{\bf b})$ and $({\bf a}'|{\bf b}')$ is
\begin{align}
({\bf a}|{\bf b})\odot ({\bf a}'|{\bf b}') & = ({\bf a}|{\bf b})\left[\begin{array}{cc} 0 & I\\ I & 0\end{array}\right] ({\bf a}'|{\bf b}')^T \\
& = {\bf a}\cdot {\bf b}'\oplus {\bf a}'\cdot {\bf b}. 
\end{align}
The symplectic weight of a vector $({\bf a}|{\bf b})$ is the number of indices $i$ at which either $a_i$ or
$b_i$ is nonzero. $E_{({\bf a}|{\bf b})}$ is defined by $e_1\otimes e_2\otimes \dots \otimes e_n$ 
where $e_i$ equals $I$ if $(a_i,b_i)=(0,0)$, $X$ if $(a_i,b_i)=(1,0)$,
$Z$ if $(a_i,b_i)=(0,1)$, and $Y$ if $(a_i,b_i)=(1,1)$ and the associated projector is
$P_{({\bf a}|{\bf b})}=\frac{1}{2}(I+E_{({\bf a}|{\bf b})})$.

The next definition specifies the ingredients of an AC06 quantum error-correcting code (AC06 QECC).
Theorem 1 of \cite{Calderbank:06a} defines a quantum code, but our definition of an AC06 QECC is based instead
on Theorem 2 of \cite{Calderbank:06a}, which provides sufficient conditions for the code
to be an error-correcting code.

\begin{definition}[AC06 QECC]
Let $f$ be an $n$ variable Boolean function and let $x_1,x_2,\dots,x_{2n}$
be a list of the $n$-bit column vectors of an $n\times 2n$ matrix $A_f$. An AC06 QECC
with data $(f,\{x_i\}_{i=1}^{2n})$ is the image of the projector $f(P_1,P_2,\dots,P_n)$,
where (i) the rows of $A_f$ are linearly independent with pairwise symplectic inner product zero
and (ii) $P_i=P_{({\bf a}_i|{\bf b}_i)}$ is associated to the $i$th row of $A_f$.
\end{definition}

\begin{theorem}[Theorem 2 of \cite{Calderbank:06a}]
Let $D_d$ be the set of all $2n$-bit vectors of symplectic weight less than $d$.
An AC06 QECC with data $(f,\{x_i\}_{i=1}^{2n})$ is an $((n,K,d))$ quantum code if
$f$ has weight $K$ and $\{ A_fw^T\ |\ w\in D_d \}\subseteq Cset_f$.
\end{theorem}

The main result of this subsection, stated and proven next, is that AC06 QECCs are CWS codes.

\begin{theorem}\label{thm:AC06eqCWS}
An AC06 quantum error-correcting code is a codeword stabilized quantum code.
\end{theorem}

\begin{IEEEproof}
Consider an AC06 QECC with data $(f,\{x_i\}_{i=1}^{2n})$. The matrix $A_f$, whose $2n$ columns are
$\{x_i\}_{i=1}^{2n}$, has linearly independent rows with pairwise symplectic inner products that are zero. 
Therefore, $A_f$ corresponds naturally to a group generated by $n$ pairwise commuting operators $\{g_i\}_{i=1}^n$ 
from the $n$ qubit Pauli group. Let $|S_{\bf c}\rangle$ be the state stabilized by 
$S=\langle (-1)^{c_i}g_i\rangle_{i=1}^n$ for some $n$-bit vector ${\bf c}$. 
A nonzero Boolean function $f$ can be represented as 
\beq
f({\bf v})=\sum_{{\bf c}\in {\cal C}_f} v_1^{c_1}v_2^{c_2}\dots v_n^{c_n},
\eeq
which corresponds, in this case, to the projector
\beq
f(P_1,P_2,\dots,P_n) = \sum_{{\bf c}\in {\cal C}_f} P_1^{c_1}P_2^{c_2}\dots P_n^{c_n},
\eeq
where $P_i^0=\bar{P_i}=\frac{1}{2}(I-g_i)$ and $P_i^1=P_i=\frac{1}{2}(I+g_i)$.
The term $P_1^{c_1}P_2^{c_2}\dots P_n^{c_n}$ projects onto the state $|S_{\bar{{\bf c}}}\rangle$,
where $\bar{\bf c}=\bar{c_1}\bar{c_2}\dots\bar{c_n}$, therefore
\beq
f(P_1,P_2,\dots,P_n) = \sum_{{\bf c}\in \bar{\cal C}_f} |S_{{\bf c}}\rangle\langle S_{{\bf c}}|.
\eeq
Hence, the AC06 QECC is spanned by a set of eigenstates of a stabilizer $S$, each of which
has a vector of eigenvalues given by a codeword ${\bf b}$ in the inverted code 
$\bar{\cal C}_f$, where $b_i=0$ indicates a $+1$ eigenvalue for $g_i$ and $b_i=1$ indicates a $-1$ eigenvalue
for $g_i$.
To establish correspondence with a CWS code, we need to show that there is a mapping
$W$ from $n$-bit strings ${\bf c}$ to Pauli operators $W({\bf c})$ such that 
$|S_{\mathbf{c}}\rangle=W(\mathbf{c})|S_{\bf 00\dots 0}\rangle$. Indeed, there is a
Clifford circuit $U$ that encodes $U|\underbrace{00\dots 0}_n\rangle=|S_{\bf 00\dots 0}\rangle$
and acts like $UZ_iU^\dag=g_i$ for $i=1,\dots,n$. Therefore,
$UX_iU^\dag$ anticommutes with $g_i$ and commutes with all $g_j$, $j\neq i$.
By this observation, the map
\begin{equation}
W(\mathbf{c}):=\prod_{i=1}^n \left[ UX_iU^\dag \right]^{c_i}
\end{equation}
has the desired properties, and we obtain the set of CWS word operators 
$W(\bar{\mathcal C}_f)$ by applying $W$ to each codeword in $\bar{\mathcal C}_f$.
Therefore, the AC06 QECC with data $(f,\{x_i\}_{i=1}^{2n})$ is associated with
a CWS code (not in standard form) with stabilizer state $|S\rangle$ corresponding to $A_f$, classical
code $\bar{\mathcal C}_f$, and word operators $W(\bar{\mathcal C}_f)$.
\end{IEEEproof}

The mapping can be inverted to obtain data for an AC06 QECC from a CWS code 
as well. There is freedom in the choice of generating set for the stabilizer
state in the CWS construction so it may be necessary to conjugate by a Pauli 
operator to fix the signs of the stabilizer generators to $+1$ before mapping
them to the column vectors $\{x_i\}_{i=1}^{2n}$.

\begin{example}\label{ex:ac06tocws}
This detailed example demonstrates the mapping given in the proof of Theorem~\ref{thm:AC06eqCWS}
from an AC06 QECC $(f,\{x_i\}_{i=1}^{2n})=(f,A_f)$ to a CWS code $(S_A,{\mathcal C}',W(\bar{\mathcal C}_f))$.
The AC06 $((5,6,2))$ code is given by the boolean function
\begin{align*}
f(v) & = v_1v_2v_3 + v_3v_4v_5 + v_2v_3v_4 \\
& + v_1v_2v_5 + v_1v_4v_5 + v_2v_3v_4v_5
\end{align*}
and the matrix
\begin{equation*}
A_f = \left[\begin{array}{cccccccccc}
0 & 0 & 1 & 1 & 0 & 0 & 1 & 1 & 1 & 1 \\
0 & 1 & 1 & 0 & 0 & 1 & 1 & 1 & 1 & 0 \\
1 & 1 & 0 & 0 & 0 & 1 & 1 & 1 & 0 & 1 \\
1 & 0 & 0 & 0 & 1 & 1 & 1 & 0 & 1 & 1 \\
0 & 0 & 0 & 1 & 1 & 0 & 1 & 0 & 0 & 0 \end{array}\right].
\end{equation*}
First, consider the boolean function $f$.
Indeed, $f(v)$ is a function of $n=5$ variables and has weight $K=6$. This can
be seen by writing $f$ in the form
\begin{align*}
f(v) & = \sum_{\mathbf{c}\in \{0,1\}^n} f(\mathbf{c})v_1^{c_1}\dots v_n^{c_n}
= \sum_{\mathbf{c}\in {\mathcal C}_f} v_1^{c_1}\dots v_n^{c_n} \\
& = v_1v_2v_3\bar{v_4}\bar{v_5}+\bar{v_1}\bar{v_2}v_3v_4v_5+\bar{v_1}v_2v_3v_4\bar{v_5}\\
& +v_1v_2\bar{v_3}\bar{v_4}v_5+v_1\bar{v_2}\bar{v_3}v_4v_5+\bar{v_1}v_2v_3v_4v_5
\end{align*}
where $v_i^{c_i}$ equals $v_i$ if $c_i=1$ and $\bar{v_i}$ if $c_i=0$.
The classical code ${\mathcal C}_f$ is the set of $n$-bit strings on which $f$
evaluates to $1$, i.e. $11100$, $00111$, $01110$, $11001$, $10011$, and
$01111$.
Second, observe that the rows of $A_f$ are indeed linearly independent and
pairwise orthogonal in the symplectic inner product. The rows of $A_f$
correspond to stabilizer generators $E_1=IZYYZ$, $E_2=ZYYZI$, $E_3=YYZIZ$, 
$E_4=YZIZY$, and $E_5=IZIXX$, respectively. These are the generators of the
stabilizer $S_A$ for the state $|S\rangle$. The AC06 construction uses the fact 
that the projectors $P_y=\frac{1}{2}(I+E_y)$, $y=1,\dots,n$, are pairwise
commutative projection operators of dimension $2^{n-1}$ and 
$P_1P_2\dots P_n$, $P_1P_2\dots \tilde{P_n}$, \dots, 
$\tilde{P_1}\tilde{P_2}\dots\tilde{P_n}$ are all $1$-dimensional projection
operators, so that $P_f:=f(P_1,\dots,P_n)$ is a projector onto a subspace
of dimension $wt(f)$ (Theorem 1 of \cite{Calderbank:06a}), where the boolean operations 
are replaced by the operations defined in Definition 6 of 
\cite{Calderbank:06a}. Considering just the first term of $P_f$, we see that
\begin{align*}
P_1\wedge & P_2\wedge P_3\wedge \tilde{P_4}\wedge\tilde{P_5} \\
& = P_1P_2P_3(I-P_4)(I-P_5) \\
& = \frac{1}{2^5} (I+E_1)(I+E_2)(I+E_3)(I-E_4)(I-E_5)
\end{align*}
is a projector onto a stabilizer state $W_1|S\rangle$ where $W_1$ is a
Pauli operator that commutes with $\{E_1,E_2,E_3\}$ and anticommutes with
$\{E_4,E_5\}$, i.e. $W_1=Z_5$. Notice that the partition of the generators
into commuting and anticommuting sets is given by the first codeword
$11100$ of ${\mathcal C}_f$. The terms are combined using 
the operation $P\oplus Q=P+Q-2PQ$, which equals $P+Q$ when the projectors 
are pairwise orthogonal, as they are when $P$ and $Q$ project onto stabilizer 
states. Therefore, $P_f=\sum_{i=1}^K W_i|S\rangle\langle S|W_i^\dag$ where 
the $W_i$ are chosen to commute or anticommute with the generators of the
stabilizer of $|S\rangle$ according to the codewords of ${\mathcal C}_f$. We
conclude that the AC06 $((5,6,2))$ code is a CWS code with stabilizer
$\langle IZYYZ, ZYYZI, YYZIZ, YZIZY, IZIXX\rangle$ and word operators
$\{ Z_5,Z_3,Z_4,Z_1,Z_2,X_3X_4X_5\}$ that correspond to the classical code 
${\mathcal C}'=\bar{\mathcal C}_f=\{00011,11000,10001,00110,01100,10000\}$
specifying the generator's signs for each basis state of the quantum code.
We can arrange for the all-zeros codeword to be in ${\mathcal C'}$ by
multiplying each word operator by $X_3X_4X_5$ (and, hence, adding $10000$
to each codeword in ${\mathcal C}'$). This is a local operation, so the
code parameters do not change.
\end{example}

\subsection{Mapping from AC06 to the standard form of CWS}

Three distinct steps may be identified, in building a mapping between
the AC06 $(A_f,f)$ code, and the CWS $(\mathcal{G},\mathcal{C})$ code
in standard form,
\begin{equation}
(A_f,f) \stackrel{Stab}{\longrightarrow} (S_A,{\mathcal C}') %
	\stackrel{LC}{\longrightarrow} ({\mathcal G_A},{\mathcal C}') %
	\stackrel{Gen}{\longrightarrow} ({\mathcal G},{\mathcal C}) %
\,.
\end{equation}

First, $(A_f,f)$ is re-written as a stabilizer $S_A$ and a classical
code ${\mathcal C}'$, using standard definitions.  The subscript $A$
on $S_A$ reminds us that the stabilizer is generated by the generators
$g_A=\langle g_1,\ldots,g_n\rangle$, where each generator $g_k$
corresponds to a row of $A_f$. Second, a (non-unique) local Clifford
transform $L$ turns $S_A$ into $\mathcal{G}_A$, leaving ${\mathcal
C}'$ invariant.  $\mathcal{G}_A$ is a graph state with generators
$Lg_AL^{\dagger}$.  Third, careful choice of appropriate generators
turn the classical code ${\mathcal C}'$ into the ${\mathcal C}$ used
in the CWS construction.  A fourth issue that arises is the
limitation on $f$ needed to allow degenerate codes to be considered.
These three steps and the degeneracy issue are discussed below, one at
a time.

\subsubsection{$(A_f,f) \stackrel{Stab}{\longrightarrow} (S_A,C')$}

We have already accomplished this step by way of Theorem~\ref{thm:AC06eqCWS},
but we review it quickly to show the entire chain of steps to achieve
standard form.
The $n\times 2n$ matrix $A_f$ describes the generators of a quantum
stabilizer state, which we may denote as $S_A$, when the left $n\times
n$ half is interpreted as describing $X$ Pauli terms, and the right
half, $Z$ Pauli terms, following the standard
prescription\cite{Nielsen00b}.  Let the generators of this stabilizer
be $g_A=\langle g_1,\ldots,g_n\rangle$; each generator $g_k$
corresponds to a row of $A_f$.  Let $|S\>$ be the quantum state
stabilized by $S_A$.

\def\>{\rangle}
\def\<{\langle}

The Boolean function $f$ defines a classical code, through its action
on the $K$ bit strings $\mathbf{c}'_j = j_1\ldots j_n$; explicitly,
we may define
\begin{equation}
	\mathcal{C}'=\{ \mathbf{c}'_j|f({\bar{\mathbf c}'_j})=1\}
\,,
\label{Cprime}
\end{equation} 
where ${\bar{\mathbf c}'_j}$ denotes the complement of ${{\mathbf
c}'_j}$ (needed because of how $f$ is defined in AC06, see 
Example~\ref{ex:ac06tocws}).

In the CWS standard form, the all-zeros codeword is in the classical code
${\mathcal C}'$, i.e. the state $|S\rangle$ is in the code. This can be
arranged by choosing one of the states $|S_{\mathbf{c}'_j}\>$ in the
code and applying to the whole code the local Pauli operation that maps 
$|S_{\mathbf{c}'_j}\>$ to $|S\>$. Since this has no effect on the stabilizer
$S_A$, and the resulting code is locally equivalent to the original code,
we now assume without loss of generality that ${\mathcal C}'$ 
contains the all-zeros codeword.

\subsubsection{$(S_A,{\mathcal C}') \stackrel{LC}{\longrightarrow} 
		({\mathcal G}_A,{\mathcal C}')$}

The second step needed is an intermediate, but simple map,
transforming $S_A$ into graph state form\cite{VanDenNest:04a}. This
can be done using Clifford operations on individual qubits (``LC
transformations'').  Importantly, though, we must also keep track of
how ${\mathcal C}'$ transforms when the stabilizer $S_A$ is
transformed, since ${\mathcal C}'$ is partially defined in terms of
$S_A$.

Let $L=\bigotimes\limits_{i=1}^n L_i$ be the $n$-qubit operation given
by the tensor product of single qubit Clifford operations $L_i$.
When transformed by $L$, the generators of the stabilizer $S_A$ map to
become 
\begin{equation}
	\< g_1,...,g_n\> \rightarrow \< g'_1,...,g'_n\>
\,,
\label{LC}
\end{equation} 
where $g'_i=Lg_iL^{\dagger}$.  Since $L$ also transforms $w_j$ to
$w'_j=Lw_jL^\dagger$, it follows that the commutation relations of
$w'_j$ with $g'_k$ are the same as between $w_j$ and $g_k$.  Thus, LC
transformations leave ${\mathcal C}'$ unchanged, mapping
$(S_A,{\mathcal C}')$ into $({\mathcal G}_A,{\mathcal C}')$.  Again,
just as for $S_A$, the subscript $A$ on ${\mathcal G}_A$ reminds us
that the generator of this graph state is $Lg_AL^{\dagger}$, and
originates from $A_f$.

\subsubsection{$({\mathcal G}_A,{\mathcal C}') \stackrel{Gen}{\longrightarrow} 
		({\mathcal G},{\mathcal C})$}

The final step in transforming the quantum code into CWS form involves
nailing down a degree of freedom which allows ${\mathcal C}$ to be
changed, without changing the stabilizer, or the quantum code
specified.  In particular, $\mathcal{C}'$ is dependent on the choice
of generators for ${\mathcal G}_A$.  Let $R$ be a binary valued,
invertible $n\times n$ matrix $R_{ji}$, which transforms a generator
set $\< g_1,g_2,\ldots,g_n\>$ into $\< g'_1,g'_2,\ldots,g'_n\>$, where
\begin{equation}
	g'_i= \prod_{j=1}^n g_j^{R_{ji}}
\,.
\end{equation}
We may keep track of this transform by rewriting ${\mathcal G}_A$ as
${\mathcal G}$, though, of course, the stabilizer (and thus the
corresponding graph) must be left unchanged when the generator set is
changed.  Upon this transformation by $R$, the code ${\mathcal C}'$
must also be transformed, to keep the quantum code invariant.
Specifically, if $\mathcal{C}'$ is written as a $K\times n$ matrix,
then:
\begin{theorem}
The quantum code $({\mathcal G}_A,\mathcal{C}')$ is the same as the
quantum code $({\mathcal G},\mathcal{C}'R)$.  That is, if the
stabilizer generators are changed by $R$, the code must also be
transformed by matrix multiplication by $R$.
\end{theorem}
\begin{IEEEproof} 
We have $w_jg_kw_j=(-1)^{j_k}g_k$, and we want to calculate $j_k'$ given
by $w_jg_k'w_j=(-1)^{j_k'}g_k'$. Note 
\begin{align*}
w_jg_k'w_j & = w_j\prod_{k=1}^n g_k^{R_{kt}}w_j = \prod_{k=1}^n w_jg_k^{R_{kt}}w_j \\
& =\prod_{k=1}^n (w_jg_kw_j)^{R_{kt}}= \prod_{k=1}^n ((-1)^{j_k}g_k)^{R_{kt}} \\
& =\prod_{k=1}^n ((-1)^{j_kR_{kt}}g_k^{R_{kt}})=(\prod_{k=1}^n (-1)^{j_kR_{kt}})(\prod_{k=1}^n g_k^{R_{kt}}) \\
& = ((-1)^{\oplus_{k=1}^n j_kR_{kt}})\prod_{k=1}^n g_k^{R_{kt}}= (-1)^{j_k'}g_k',
\end{align*}
which gives $j_k'=\oplus_{k=1}^n j_kR_{kt}$.
\end{IEEEproof}

Essentially, this equivalence indicates that row reductions in the
symplectic $n\times 2n$ form of the stabilizer can leave the quantum
code invariant, if the same row reduction is done to the binary code.
Moreover, LC equivalence and the choice of generators of the
graph state do not change the error correcting property of the quantum
code.  Thus, using a row reduction transform $R$, and letting ${\mathcal
C} = {\mathcal C'}R$, we conclude that $({\mathcal G},{\mathcal C})$
is a CWS code with dimension and distance identical to the original
AC06 code $(A_f,f)$.

It must be noted that the row reduction does change the errors (in
terms of binary strings) detected by the classical code.  More
precisely, for a CWS code $({\mathcal G},{\mathcal C})$ in the
standard form that we have obtained from an AC06 code $(A_f,f)$, 
we may define a corresponding $(A'_{f'},f')$ in the
language of AC06, by
\begin{eqnarray}
  f'(\bar{\mathbf c}_j)&=& 1,\ \forall\ {\mathbf c}_j\in \mathcal{C}
\\	A'_{f'} &=& [I \, \Lambda]
\,,
\end{eqnarray}
where $I$ is the $n\times n$ identity matrix, and $\Lambda$ is the
adjacency matrix of the graph $\mathcal{G}$.

The complementary set $Cset_{f'}$ of the Boolean function $f'$ is no longer the same
as the the complementary set $Cset_f$ of the Boolean function $f$, but they have same size
due to the linearity of the transform relating $\mathcal{C'}$ and
$\mathcal{C}$. Moreover, given quantum code distance $d$, the set of
induced classical error strings $Cl_{\mathcal{G}}({\cal E})$ for
$({\mathcal G},{\mathcal C})$ is indeed the AC06 error set, specified as
$\{x_1,x_2\ldots x_{2k}\}*w^T$ in Theorem~2 of \cite{Calderbank:06a},
a subset of the complementary set $Cset_{f'}$ of $f'$.

\subsubsection{Degenerate codes}

The AC06 framework does not discuss how to allow for
degenerate quantum codes, whereas the CWS construction includes these
explicitly.  The above mapping of AC06 to the standard form CWS codes
applies only to non-degenerate codes, but the method indicates how
degenerate codes can also be constructed using the AC06 framework, as
follows.  Specifically, one must appropriately constrain the Boolean
function $f$ (ie $\mathcal{C}'$).

All degenerate quantum codes can be expressed using a certain form for
${\mathcal C}'$, illustrated by the following.  Consider a degenerate
code of distance $d$, given stabilizer $S$.  Define the set
\begin{eqnarray}
	S_d &=& \{E | E\in S ~{\rm and}~\text{wt}(E)<d \} 
\nonumber \\
	&& \cup\  \{-E | E\in -S ~{\rm and}~\text{wt}(E)<d \} 
\,,
\end{eqnarray}
where $\text{wt}(E)$ gives the weight of the Pauli operator $E$. If
the rank of $S_d$ is $r$, then $r$ independent elements $g_1,\ldots
g_r\in S_d$ can be chosen, such that
$\<g_1,\ldots,g_r,g_{r+1},\ldots,g_n\>$ generate $S$, but
$g_{r+1},\ldots g_n$ are not in $S_d$.  According to the CWS
construction described in the first step above, these generators imply
a representation of a classical code $\mathcal{C'}$ with each codeword
being $0$ for the first $r$ coordinates. In other words, $\<
g_1,\ldots ,g_r\>$ stabilizes $(A_f,f)$.  Due to the one--to--one
correspondence between $f$ and $\mathcal{C}'$, this gives a structure
for the values of $f$, from which a search for degenerate codes can
initiate.

\subsection{The algorithm $\&$ complexity}

Given the equivalence between AC06 and CWS codes, it is insightful to
compare the algorithms implied by each for finding new codes.  Both
approaches construct a quantum code $(\mathcal{G},\mathcal{C})$, but
each analyze and calculate from different starting points.  The search
algorithm based on the CWS construction starts from the analysis of
the structure of a given $\mathcal{G}$, takes a specification the
desired properties of $\mathcal{C}$, and searches for a satisfactory
$\mathcal{C}$, eg using the maximum clique algorithm.  In contrast,
the search algorithm based on the AC06 framework starts from the
analysis of the structure of a given $f$ (ie, $\mathcal{C}'$), and
searches for a stabilizer state $A_f$ which is LC equivalent to some
graph state $\mathcal{G}$.  This is why the two methods are in a
sense, the mirror image of each other.

How do the computational complexities of the two approaches compare?
AC06 implies an algorithm starting from a given classical code $f$ to
find the quantum code $(A_f,f)$.  This suggests a need to consider
$2^{2^n}$ different Boolean functions.  In contrast, the {\sc
cws-maxclique} algorithm starts from $2^{n \choose 2}$ possible
graphs (or ideally, a smaller set of just the different ones).

However, this comparison is incomplete. In practice, if we really want
to find an particular $((n,K,d))$ code, then there will be
${2^n\choose K}$ classical codes to look at, and for each code the
AC06 algorithm needs to search for $\sim 2^{2n^2}$ possible sets of
strings. For a given classical code, to check whether a particular
string is in the complementary set $Cset_f$ of the code takes $K^2$ steps. And to check
whether a chosen set of $2n$ strings gives a valid stabilizer state
$[A\, B]$ needs $n^2$ steps. Therefore, with the AC06 algorithm, the
complexity of searching for an $((n,K,d))$ code is roughly
\begin{equation}
	n^2 K^2 2^{2n^2}{2^n \choose K}
\,.
\end{equation}
This is comparable but slightly worse than the result obtained for the {\sc
cws-maxclique} algorithm, in Eq.~(\ref{complexity}).

Some simplifications used in {\sc cws-maxclique} may also apply to
AC06; in particular, a reduction of the code search space due to LC
invariance should be considered.  In practice, in order to find all
quantum codes $(A_f,f)$, we only need to consider the codes
$\mathcal{C}'$ equivalent under column reductions. For $K\geq n$, this
LC equivalence is the same as equivalence classification of all the
$((K,n'))$ binary linear codes, where $n'\leq n$.  For fixed $n'$, the
number of such codes is given by the Gaussian binomial factor ${2^K
\choose n'}_{Gaussian}$ \cite{MacWilliams:77}. Note this
classification gives not only all the $((n',K))$ codes $\mathcal{C}'$
we need to start with, but also all the $((n',K'\leq K))$ codes
$\mathcal{C}'$. For instance, the $((K=4,n'=3))$ code
$\{(0,0,0,0),(0,0,0,1),(0,0,1,0)\}$, viewed by column, is an
$((n'=3,K'=3))$ code $\{(0,0,0),(0,0,1),(0,1,0)\}$, but not an
$((n'=3,K=4))$ code.

\section{The structure theorems}
\label{sec:struct}

The ability to search for CWS codes through solving the {\sc
maxclique} problem is unsurprising; any unstructured search problem
can be reduced to an NP-complete problem.  Thus, as it stands, the
{\sc cws-maxclique} algorithm presented in Section~\ref{sec:alg} is
unsatisfactory (at least, for large cases), for the search space grows
exponentially with the problem size $n$.  Moreover, as shown in
Section~\ref{sec:bfunc}, the complexity of the AC06 algorithm is
comparably bad, and is thus also unsatisfactory.

Since a major goal of the study of nonadditive codes is identification
of codes with parameters superior to all possible additive codes,
pruning the search space is worthwhile as a first step, before
applying such brute-force search.

Is there hope?  All nonadditive quantum codes with good parameters
constructed so far have been CWS codes, as was shown in
\cite{CSSZ:07}. Also, very recently the $((10,24,3))$ CWS code was
enumerated\cite{Yu:07b}; this code saturates the linear programing
bound on code parameters.  It thus seems that we should be optimistic
about finding more CWS codes that outperform additive codes.
We call an $((n,K,d))$ additive quantum code {\it optimal} if there does not
exist any $((n,2K,d))$ additive quantum code.
One might hope that improved codes could be built from optimal $((n,K,d))$
additive codes, using the idea that these
codes could be subcodes of larger (non-additive) CWS codes with
superior parameters. If this were true, then a promising strategy
would be to start with the optimal additive codes and try to increase
the dimension.

This strategy leads to useful knowledge about the structural
properties of CWS codes and reveals relations between codes with
parameters $((n,K,d))$ and $((n,K',d))$, where $K'>K$. These relations are
especially interesting when given extra knowledge about the nature of
the classical code ${\mathcal C}$ employed in the construction.
Surprisingly, we find that the low-dimensional CWS codes are
actually additive. In particular, we find that all $((n,3,d))$ CWS
codes are subcodes of some $((n,4,d))$ additive codes.  Furthermore, we
find restrictions on how optimal additive codes can and cannot be
subcodes of larger CWS codes.  

Before presenting these structure theorems, we review the 
relationship between the linearity of $\mathcal{C}$ and the additivity of $\mathcal{Q}=(\mathcal{G},\mathcal{C})$.

\subsection{Linearity of $\mathcal{C}$ and additivity of $\mathcal{Q}=(\mathcal{G},\mathcal{C})$}

Recall from Theorems~4 and 5 in \cite{CSSZ:07} that the following 
facts are true:

\begin{fact}
\label{fact1}
If $\mathcal{C}$ is a linear code (or equivalently, the word operators form a group),
then $\mathcal{Q}=(\mathcal{G},\mathcal{C})$ is an additive code.
\end{fact}

\begin{fact}
\label{fact2}
If $\mathcal{Q}$ is an additive code, then there exists a linear code $\mathcal{C}$
and a graph $\mathcal{G}$, such that $\mathcal{Q}=(\mathcal{G},\mathcal{C})$.
\end{fact}

However, when ${\mathcal C}$ is nonlinear, the question of whether
$({\mathcal G},{\mathcal C})$ is additive or not is completely open,
since it may or may not
be possible that $({\mathcal G},{\mathcal C})$ is local unitary (LU) equivalent to
some additive code.

The following example explicitly illustrates this possibility, by
presenting two CWS codes: $({\mathcal G},{\mathcal C}_2)$ with
nonlinear ${\mathcal C}_2$, and $({\mathcal G},{\mathcal C}_1)$ with
linear ${\mathcal C}_1$.  The two codes are LU equivalent to each
other:
\begin{example} 
\label{ex:lucode}
Let
\begin{eqnarray}
	\mathcal{G}&=&\<XZZZ,ZXII,ZIXI,ZIIX\>
\\	\mathcal{C}_1 &=& \{0000, 0110, 0101, 0011 \}
\\	\mathcal{C}_2 &=& \{0000, 0110, 0101, 1011 \}
\,. 
\end{eqnarray}
Note that $(\mathcal{G},\mathcal{C}_1)$ is an additive code since the
codewords of $\mathcal{C}_1$ form a group under binary addition (it is
thus a linear code).  In contrast, since $\mathcal{C}_2$ is nonlinear
(its set of codewords are not closed under addition),
$(\mathcal{G},\mathcal{C}_2)$ is not LC equivalent to any additive
code.  Nevertheless, we can show that ${\mathcal Q}_1 =
(\mathcal{G},\mathcal{C}_2)$ is LU equivalent to ${\mathcal Q}_2 =
(\mathcal{G},\mathcal{C}_1)$, by giving an explicit LU equivalence
between the projectors into the two quantum code spaces, $P_{1}$ and
$P_{2}$.  For this purpose, it is convenient to first transform by
$H_{234}=H_2\otimes H_3\otimes H_4$ and disregard normalization factors, 
such that
\begin{eqnarray} 
	 	P'_{1}&=& H_{234}P_{1}H_{234}
\nonumber\\   	      &=& I+XXXX+YYYY+ZZZZ 
\\	 	P'_{2} &=& H_{234}P_{2}H_{234}
\nonumber\\ 		&=& I+ZZZZ
\nonumber\\ 		&& +\frac{1}{2}(XXXX+YYYY+XXYY+YYXX
\nonumber\\ 		&& -XYYX-YXXY-XYXY-YXYX) 
\,.
\end{eqnarray} 
From Theorem~4.2 of \cite{Roychowdhury:97}, LU equivalence need only
consider $U=U_1\otimes U_2\otimes U_3\otimes U_4$ where $U_i$ maps $X$
to $aX+bY$ and $Y$ to $bX-aY$.  We find that
$UP'_{1}U^{\dagger}=P'_{2}$, if $U$ is defined such that 
\begin{eqnarray}
    U_i X_i U_i^\dagger &=& [ X_i - (-1)^{\lfloor i/2\rfloor} Y_i ] /\sqrt{2}
\\  U_i Y_i U_i^\dagger &=& [ X_i + (-1)^{\lfloor i/2\rfloor} Y_i ] /\sqrt{2}
\,,
\end{eqnarray}
where $\lfloor i/2\rfloor$ is $0$ for $i<2$ and $1$ otherwise.  The
existence of this LU equivalence is unsurprising, since it is
known~\cite{Rains:97c} that any $((4,4,2))$ code is LU equivalent to
the additive $[[4,2,2]]$ code.
\end{example}

In general, for a CWS code $\mathcal{Q}=(\mathcal{G},\mathcal{C})$ with a 
nonlinear $\mathcal{C}$, we cannot directly infer that $\mathcal{Q}$ is 
nonadditive. However, for fixed $n$ and $d$, if we seek a code with optimal $K$ and
only find $((n,K'\geq K,d))$ codes $\mathcal{Q}=(\mathcal{G},\mathcal{C})$ with 
nonlinear $\mathcal{C}$, then we can conclude that $\mathcal{Q}$ nonadditive. 
Put another way, if we fix $n$ and $d$, do an 
exhaustive search over all the graphs and classical codes, and only find 
quantum codes with nonlinear classical codes $\mathcal{C}$ for the optimal 
$((n,K,d))$ CWS codes, then we can conclude that the optimal $((n,K,d))$ 
CWS codes we found are indeed nonadditive.
This can be shown by contradiction: 
if $\mathcal{Q}=(\mathcal{G},\mathcal{C})$ is additive, 
then there exists some local unitary operation 
$U=\bigotimes_{i=1}^{n}U_i$, where each $U_i$ is a single qubit operation, 
such that $U\mathcal{Q}U^{\dag}=\mathcal{Q}'$ and $\mathcal{Q}'$ is additive.
Then, according to Fact~\ref{fact2}, there exists a linear code 
$\mathcal{C}'$ and a graph $\mathcal{G}'$ such that 
$\mathcal{Q}'=(\mathcal{G}',\mathcal{C}')$. 

\subsection{Structure theorems}

We now present and prove some structure theorems governing
CWS codes, and provide several useful corollaries. Recall that
we say an additive $((n,K,d))$ quantum code is {\it optimal} if there is no
$((n,2K,d))$ additive quantum code.

Our first theorem concerns CWS codes with dimension $2$: 
\begin{theorem} 
	All $((n,2,d))$ CWS codes are additive. 
\label{theorem:cws_dim2}
\end{theorem} 
\begin{IEEEproof} 
By the CWS construction, an $((n,2,d))$ CWS code is spanned by basis
vectors of the form $\{w_1\ket{S},w_2\ket{S}\}$, with word operators
$w_1=I=Z^{\mathbf{c}_1},w_2=Z^{\mathbf{c}_2}$. However $\{w_1,w_2\}$
form a group. So according to Theorem~5 of \cite{CSSZ:07} (or Fact~1), this CWS
code is an additive code.
\end{IEEEproof}

A natural corollary of Theorem~\ref{theorem:cws_dim2} is
\begin{corollary} 
	If an additive code of parameters $((n,1,d))$ is optimal, then
	there do not exist any CWS codes with parameters
	$((n,K>1,d))$.
\label{Kgeq1}
\end{corollary}

From corollary~\ref{Kgeq1}, it follows that the $((7,2,3))$ and
$((9,2,3))$ nonadditive codes given in \cite{Pollatsek:03a} and the
$((11,2,3))$ code given in \cite{Roychowdhury:97} are not local
unitary (LU) equivalent to any CWS code, for they are not LU
equivalent to any additive code. This implies that there exist
codes that are outside the CWS construction, as was claimed in
Fig.~\ref{fig1}.

Now we present a theorem concerning CWS codes of dimension $3$:
\begin{theorem}
	Any $((n,3,d))$ CWS code is a subcode of some $((n,4,d))$
	stabilizer code.
\label{n3d}
\end{theorem}

\begin{IEEEproof}
By the CWS construction, any $((n,3,d))$ CWS code has the form
$(\mathcal{G},\mathcal{C}_1)$ with
$\mathcal{C}_1=\{{\mathbf c}_1\!=\!0,{\mathbf c}_2,{\mathbf c}_3\}$.  Consider a new code
$(\mathcal{G},\mathcal{C}_2)$ with
$\mathcal{C}_2=\{{\mathbf c}_1\!=\!0,{\mathbf c}_2,{\mathbf c}_3,{\mathbf c}_2 \oplus {\mathbf c}_3\}$.  From
Theorem~\ref{CSSZTheorem3}, it follows that $\mathcal{C}_1$ detects
errors in $Cl_\mathcal{G}(\mathcal{E})$.  To prove Theorem~\ref{n3d},
we need to show that $\mathcal{C}_2$ also detects those errors.  It is
clear that $\mathcal{C}_2$ is a group with generators ${\mathbf c}_2, {\mathbf c}_3$ and
that ${\mathbf c}_2 \oplus {\mathbf c}_3 \notin Cl_\mathcal{G}(\mathcal{E})$ because ${\mathbf c}_2
\oplus ({\mathbf c}_2 \oplus {\mathbf c}_3) = {\mathbf c}_3$.  Therefore $\mathcal{C}_2$ detects all
of $Cl_\mathcal{G}(\mathcal{E})$.  Theorem~\ref{CSSZTheorem3} also
requires that for each $E\in \mathcal{E}$ either $Cl_\mathcal{G}(E)\ne
0$ or for all $i$, $Z^{c_i}$ commutes with $E$.  The latter constraint
is satisfied by $\mathcal{C}_2$ since $Z^{{\mathbf c}_2 \oplus {\mathbf c}_3} E = Z^{{\mathbf c}_2}
Z^{{\mathbf c}_3} E = E Z^{{\mathbf c}_2} Z^{{\mathbf c}_3}$.  Finally, since
$\{I,Z^{{\mathbf c}_2},Z^{{\mathbf c}_3},Z^{{\mathbf c}_2\oplus {\mathbf c}_3}\}$ is a group (and thus a
linear code), according to Theorem~5 in \cite{CSSZ:07} (or Fact~1), this CWS code
is a stabilizer code.
\end{IEEEproof}

Two natural corollaries of Theorem~\ref{n3d} are:
\begin{corollary}
	If an additive code of parameters $((n,2,d))$ is optimal, then there
	do not exist any CWS codes with parameters $((n,K\!>\!2,d))$.
\end{corollary}

\begin{corollary}
	There does not exist any $((7,3,3))$ CWS code, even though the linear
	programing bound does not rule out this possibility.
\end{corollary}

The two structure theorems above imply that CWS codes with parameters
better than the optimal $((n,K,d))$ additive codes need dimension
$K\geq 4$. We do know examples where $K=4$, as the $((5,6,2))$ code
\cite{Rains:97a} and the $((5,5,2))$ code \cite{Smolin:07a} beat the
optimal additive code with parameters $((5,4,2))$
\cite{Calderbank97a}.

Theorem~\ref{n3d} says that a CWS code of dimension $3$ is a subcode
of some additive code with higher dimension.  This invites a related
question: when might an optimal additive code, of dimension $K$, be a
subcode of some CWS code of higher dimension?  Unfortunately, we can
show that in some sense, optimal additive codes cannot be subcodes of 
larger CWS codes, though we cannot show the impossibility in the most general 
setting, due to the fact that ${\mathcal C}$ may be nonlinear even if a CWS
code is additive.

Motivated by LU equivalences like the one demonstrated in 
Example~\ref{ex:lucode}, we show that if $\mathcal{C}_1$
is a linear code, then an optimal additive code
$(\mathcal{G},\mathcal{C}_1)$ cannot be a subcode of any CWS code
$(\mathcal{G},\mathcal{C}_2)$, where
$\mathcal{C}_1\subset\mathcal{C}_2$:
\begin{theorem}
  Given a CWS code $(\mathcal{G},\mathcal{C}_1)$ with parameters
  $((n,K,d))$, if $\mathcal{B}$ is a linear subcode of $\mathcal{C}$
  containing $J<K$ codewords, then there exists an additive code
  $(\mathcal{G},\mathcal{C}_2)$ with parameters $((n,K'=2J,d))$.
\label{thm:nosupercode}
\end{theorem}
\begin{IEEEproof}
By the CWS construction the classical codewords
$\mathcal{C}_1=\{{\mathbf c}_1,{\mathbf c}_2,\ldots {\mathbf c}_K\}$ of
$(\mathcal{G},\mathcal{C}_1)$ can be arranged such that $c_1=0$.  From
$\mathcal{B}$ construct the linear classical code $\mathcal{C}_2=
\{{\mathbf b}_1,{\mathbf b}_2\ldots {\mathbf b}_J,{\mathbf v} \oplus {\mathbf b}_1, {\mathbf v} \oplus {\mathbf b}_2 \ldots {\mathbf v} \oplus {\mathbf b}_J\}$
where ${\mathbf v} \in \mathcal{C}_1$ but ${\mathbf v} \notin \mathcal{B}$.  Then
$(\mathcal{G},\mathcal{C}_2)$ is clearly an $n$-qubit CWS code with
$2J$ codewords.  It is an additive (stabilizer) code by Theorem 5 of
\cite{CSSZ:07} since $\mathcal{C}_2$ is a group.

It remains to check the error-correction conditions.
Theorem~\ref{CSSZTheorem3} ensures that $\mathcal{C}_1$ detects errors
in $Cl_\mathcal{G}(\mathcal {E})$, {\em i.e.} no error can turn one
codeword into another:
\begin{equation}
	{\mathbf c}_i \oplus {\mathbf c}_j \oplus {\mathbf e} \ne 0\ {\rm for\ all\ }{\mathbf e} \in
	Cl_\mathcal{G}(\mathcal{ E})
\,.
\label{cicje}
\end{equation}
The same condition for $\mathcal{C}_2$ is 
\begin{equation}
	{\mathbf b}_i \oplus {\mathbf v}^k \oplus {\mathbf b}_j \oplus {\mathbf v}^l \oplus {\mathbf e} \ne 0
\,,
\end{equation}
where $k,l \in \{0,1\}$.  Since the $b$s are a group this reduces to
\begin{equation}
	{\mathbf b}_i \oplus {\mathbf v}^k \oplus {\mathbf e} \ne 0
\end{equation}
which is true, due to Eq.(\ref{cicje}), and the fact that ${\mathbf b}_i,0,{\mathbf v} \in
\mathcal{C}_1$ for all $i$.

Theorem~\ref{CSSZTheorem3} also tells us that for all $E\in
\mathcal{E}$ either (a) $Cl_\mathcal{G}(\mathcal{E})\ne 0$ or (b) for
all $i$, $[Z^{{\mathbf c}_i},E]=0$.  $(\mathcal{G},\mathcal{C}_2)$ has the same
graph $\mathcal{G}$ as $(\mathcal{G},\mathcal{C}_1)$ so whenever (a)
is satisfied for $(\mathcal{G},\mathcal{C}_1)$ it will be for
$(\mathcal{G},\mathcal{C}_2)$.  For $\mathcal{C}_2$ (b) becomes for
all $i=1,J$ and $k=0,1$ $[Z^{{\mathbf b}_i} Z^{{\mathbf v}^k}, E]=0$.  Again, since ${\mathbf b}_i,{\mathbf v}
\in \mathcal{C}_1$ for all $i$, this is condition is met.
\end{IEEEproof}

\begin{corollary}
  An optimal additive code $(\mathcal{G},\mathcal{C})$ (for which
  $\mathcal{C}$ must be linear) cannot be extended to become a larger
  CWS code merely by adding codewords to $\mathcal{C}$.
\label{cor:nosupercode}
\end{corollary}
\begin{IEEEproof}
If the code could be extended in this way, by adding even just one
vector, then there would exist an additive code with twice as many vectors
and the same distance as the original code.  This contradicts the
statement that the original code is optimal.
\end{IEEEproof}

These structure theorems rule out certain strategies for finding
non-additive codes with parameters superior to additive codes, but
suggest other approaches.  Since an additive $((n,K,d))$ code
$(\mathcal{G},\mathcal{C}_1)$ must have linear $\mathcal{C}_1$,
Theorem~\ref{thm:nosupercode} and corollary~\ref{cor:nosupercode} tell
us that in practice we cannot search for an $((n,K'\!>\!K,d))$ CWS code
$(\mathcal{G},\mathcal{C}_2)$ just by adding codewords to
$\mathcal{C}_1$.  However, Example~\ref{ex:lucode} hints that we may
be able to shoehorn an optimal $((n,K,d))$ additive code into a CWS
code $(\mathcal{G},\mathcal{C})$ with nonlinear $\mathcal{C}$, via
some LU transform.  This gives hope to a strategy of adding codewords
to $\mathcal{C}$ to search for $((n,K'\!>\!K,d))$ CWS codes; such hope
suggests that it is worthwhile both to further explore conditions
under which two CWS codes can be linked by an LU transform, and to
better understand the structural properties of CWS codes constructed
from nonlinear codes.

\section{Discussion}

{\sc cws-maxclique} is an algorithm which may be usefully employed in
the search for new quantum codes, both additive and non-additive, as
described by the CWS construction.  Given $n$ and $K$, the algorithm
can be used to search for an $((n,K,d))$ code
$(\mathcal{G},\mathcal{C})$, with a complexity which grows roughly as
$2^{n^2}$.  In practice, by employing a number of search space
simplifications, by pruning the set of graphs ${\mathcal G}$ to
explore based on LC equivalences, and by taking guidance from
structural theorems about CWS codes, {\sc cws-maxclique} and
randomized variants of it have been used realistically\cite{CSSZ:07}
to explore codes with parameters up to $n=11$ and $K=32$.

Many interesting questions arise in the construction of this
algorithm.  For example, it is likely that {\sc cws-maxclique} can be
improved with more memory efficient implementations; reductions to
other NP-complete problems may also allow faster exploration of
specific search spaces.  Moreover, many of the simplifications used
in {\sc cws-maxclique} should also be applicable to the algorithm
introduced by the AC06 framework; and in return, any code isomorphisms
useful in simplifying AC06 should apply to {\sc cws-maxclique}.

CWS codes present a rich structure, only partially described by the
three structural theorems presented here.  We believe that there are
promising strategies for identifying new non-additive quantum codes
based on expanding known additive codes, but such a strategy has to be
executed carefully, because of limitations imposed by the theorems.
Nevertheless, given an optimal $((n,K,d))$ additive code, there is
hope for success with a strategy of adding codewords to $\mathcal{C}$
to search for $((n,K'\!>\!K,d))$ CWS codes, because of potential LU
equivalences with some non-additive code.  This hope suggests that it
is worthwhile both to further explore conditions under which two CWS
codes can be linked by an LU transform, and to better understand the
structural properties of CWS codes constructed from nonlinear codes,
so that more new quantum codes can be found. Indeed, one successful
application of this idea results in new CWS codes encoding several 
more qubits than the best known codes \cite{Grassl:08b}. It is an open 
question to determine if these nonadditive ``quantum Goethals-Preparata 
codes'' are LU equivalent to any additive quantum code.

Finally, despite the encompassing success of the CWS construction in
describing all known non-additive codes with good parameters, we point
out that there do exist codes, such as $((7,2,3))$ and $((9,2,3))$
codes, which are outside of the CWS construction.  Since these codes
are not LU equivalent to any CWS code, further new ideas will need to
be developed to reach outside the stabilizer framework, for a complete
understanding of quantum error correction codes.

\section*{Acknowledgments}

JAS was supported by ARO contract DAAD19-01-C-0056, and AWC was
supported in part by the JST CREST Urabe Project and an internship
at the IBM T. J. Watson Research Center. We gratefully acknowledge
comments and suggestions from V. Aggarwal and A. R. Calderbank.

\bibliographystyle{IEEEtran}
\bibliography{qcodes}

\end{document}